\documentclass[10pt,twocolumn,letterpaper]{article}
\usepackage{wacv}
\usepackage{times}
\usepackage{cite}
\usepackage{float}
\usepackage{epsfig}
\usepackage{graphicx}
\usepackage{amsmath}
\usepackage{amssymb}
\usepackage{subfigure}
\usepackage{stfloats}
\usepackage{verbatim}

\wacvfinalcopy
\def\wacvPaperID{323}

\ifwacvfinal\fi
\begin{document}

\title{A GAN-based Tunable Image Compression System}

\author{Lirong Wu \\
	Zhejiang University\\
	{\tt\small wulirong@zju.edu.cn}
	\and
	Kejie Huang \\
	Zhejiang University\\
	{\tt\small huangkejie@zju.edu.cn}
	\and
	Haibin Shen \\
	Zhejiang University\\
	{\tt\small shen\_hb@zju.edu.cn}
}

\maketitle
\ifwacvfinal\thispagestyle{empty}\fi

\begin{abstract}
The method of importance map has been widely adopted in DNN-based lossy image compression to achieve bit allocation according to the importance of image contents.
However, insufficient allocation of bits in non-important regions often leads to severe distortion at low $bpp$ (bits per pixel), which hampers the development of efficient content-weighted image compression systems.
This paper rethinks content-based compression by using Generative Adversarial Network (GAN) to reconstruct the non-important regions.
Moreover, multiscale pyramid decomposition is applied to both the encoder and the discriminator to achieve global compression of high-resolution images.
A tunable compression scheme is also proposed in this paper to compress an image to any specific compression ratio without retraining the model.
The experimental results show that our proposed method improves MS-SSIM by more than 10.3\% compared to the recently reported GAN-based method\cite{agustsson2018generative} to achieve the same low $bpp$ (0.05) on the Kodak dataset.
\end{abstract}

\section{Introduction}
Efficient image compression is significant for the storage, transmission, and processing of image information.
At present, there are two types of image compression: lossy compression and lossless compression.
The key point to lossy compression is to find a balance between the compression ratio and the distortion to guarantee the image quality at low $bpp$\cite{blau2019rethinking,shannon1948mathematical}.
Recently, lossy compression based on Deep Neural Networks (DNNs) is under focused development\cite{agustsson2018generative,rippel2017real,patel2019deep,agustsson2017soft,balle2016end,galteri2017deep}.
The method of importance map has been widely adopted in DNN-based lossy image compression to achieve bit allocation according to the importance of image contents\cite{lee2018context,mentzer2018conditional}.
However, its compression performance often dramatically drops at low $bpp$. 
In addition, there seem to be few tunable DNN-based image compression methods allowing an image to be compressed to any specific $bpp$ without retraining the model.

In this paper, a novel GAN-based tunable image compression system aiming at low $bpp$ is proposed to reconstruct the non-important regions of the image to compensate for the severe distortion caused by the insufficient allocation of bits in those non-important regions.
%
%
The proposed system has been tested on the Kodak, ImageNet and Cityspace datasets. 
The experimental results show that our proposed scheme outperforms the-state-of-art schemes when $bpp$ is smaller than 0.2. 
For example, our method achieves 10.3\% higher MS-SSIM than \cite{agustsson2018generative} at low $bpp$ (0.05) on the Kodak dataset.
Moreover, our method can compress images to specified compression ratios without retraining the model.
In contrast, the compression ratio of an image is unchangeable in \cite{li2018learning,mentzer2018conditional} because the importance map is deterministic for a given network structure.  Therefore, they have to modify and retrain the model to generate new importance maps. 
Our contributions are listed as follows:

\begin{itemize}
	\item We rethink content-based image compression under the GAN setting to reconstruct the non-important regions. We find that insufficient allocation of bits in non-important regions greatly limits the performance of content-based compression algorithms at low $bpp$.
	\item Unlike other methods using multiple complex networks to generate semantic maps and masks\cite{agustsson2018generative}, we design a simple network (Masking) to identify the important regions of the image and generate the importance map to guide the allocation of bits.
	\item Different from \cite{rippel2017real}, we use the multiscale structure not only in the encoder but also in the discriminator. The symmetrical multiscale structure makes it more adaptable to different sizes of objects at both the encoding end and the decoding end.
	\item We introduce tunability into our system. Unlike \cite{li2018learning,mentzer2018conditional}, we achieve different compression ratios through an user-defined parameter $n$ without retraining the model. 
\end{itemize}

The rest of the paper is organized as follows: 
In Section~\ref{sec 2}, some common image compression algorithms and techniques are briefly reviewed.
Section~\ref{sec 3} describes the entire architecture and loss function of our model.
Section~\ref{sec 4} presents our experimental results and comparison with other methods.
Section~\ref{sec 5} analyzes and summarizes our results and Sectio~\ref{sec 6} draws the conclusion.

\section{Related Work} \label{sec 2}
Recently, image compression based on deep learning has been a hot research topic.
Up to now, data autoencoder\cite{agustsson2017soft,balle2016end,zhou2018variational,theis2017lossy,balle2018variational,minnen2018joint} and Recurrent Neural Networks (RNNs)\cite{toderici2015variable,toderici2017full} are the two widely used models in the image compression architecture.
Early works using block compression decompose the image into blocks, which are then compressed and composited\cite{li2018learning,mentzer2018conditional}. Recently, global compression of the entire high-resolution image is attracting more and more attention\cite{agustsson2018generative,rippel2017real,toderici2017full,wang2018high}.
%
%

GAN has been hailed as one of the greatest achievements in the field of deep learning in recent years. 
The idea is to construct a generator and a discriminator\cite{goodfellow2014generative}.
The training purpose of the discriminator $D(\cdot)$ is to maximize its discriminative accuracy, and the training goal of the generator $G(\cdot)$ is to improve the authenticity of its reconstructed image as much as possible.
In the training process, GAN adopts an alternating optimization method, and its objective function can be expressed by the following formula:
\begin{equation}
\min\limits_{G}\min\limits_{D}\ \mathbb{E}[logD(x)]
+\mathbb{E}\big[log(1-D(G(x)))\big]
\end{equation}
With the emergence of all kinds of variants such as conditional GAN\cite{mirza2014conditional} and CycleGAN\cite{zhu2017unpaired}, GANs have been widely applied in the field of computer vision\cite{agustsson2018generative,rippel2017real,zhu2017unpaired,zhu2016generative}.
In the beginning, GAN is difficult to generate high-resolution images, which greatly limits its application.
Recently, GAN is under intense development, and the high-resolution images can be synthesized by GAN\cite{wang2018high,brock2018large}. 
For example, TC Wang et al. present a method for synthesizing 2048$\times$1024px
photo-realistic images from semantic label maps
using conditional
GAN in \cite{wang2018high}. 
Therefore, GAN is adopted to achieve global image compression\cite{agustsson2018generative}. 

\begin{figure}[h]
	\begin{center}
		\includegraphics[width=1\linewidth]{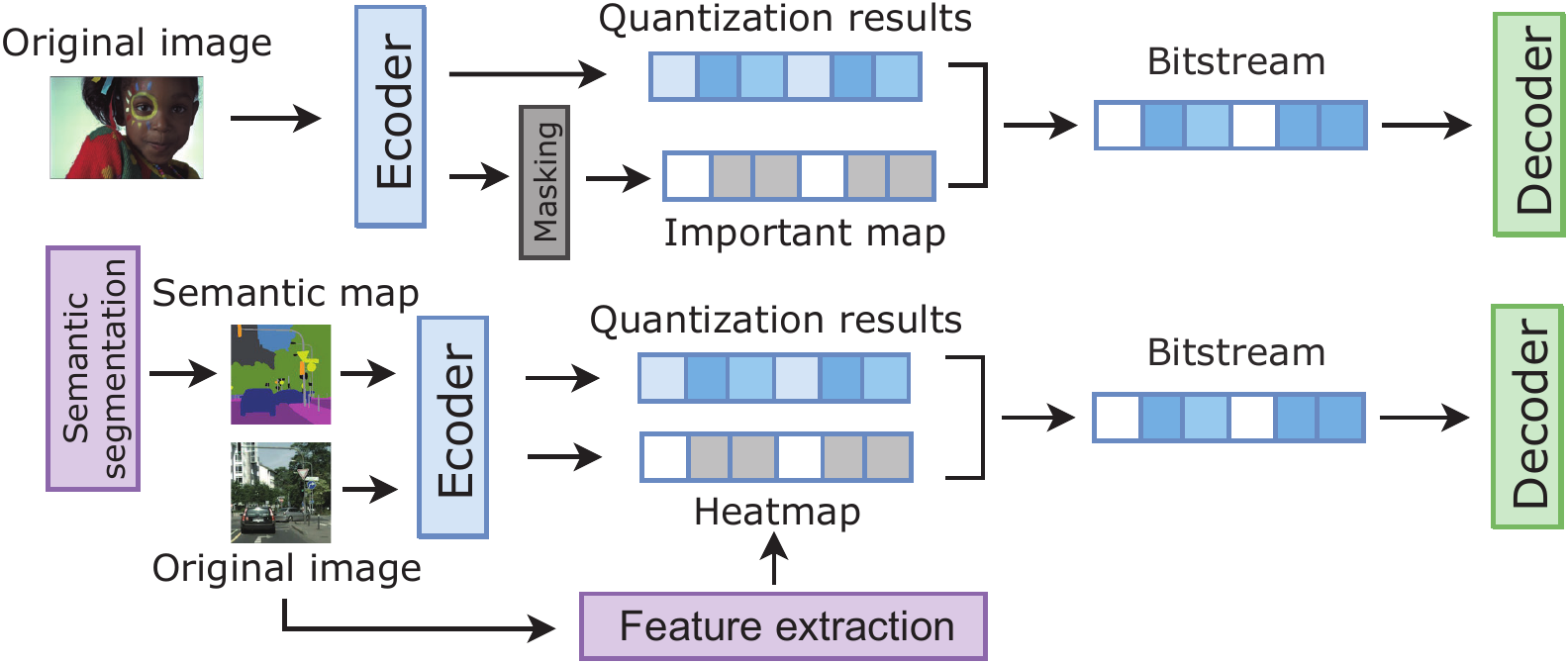}
	\end{center}
	\caption{Top: our method, bottom: other method \cite{agustsson2018generative}}
	\label{fig:com_1}
\end{figure}

At present, some GAN-based image compression methods have been proposed\cite{agustsson2018generative,rippel2017real,santurkar2018generative,galteri2017deep}, but neither of them considers the influence of image content importance on bit allocation, which limits GAN's effect on image compression.
The method proposed by Santurkar et al. is to train thumbnail images to get an efficient generator, but the information contents of the thumbnail image are so low that GAN can't play much of a role\cite{santurkar2018generative}.
The closest work to us is \cite{agustsson2018generative}, which trains a GAN-based system to achieve the compression of images.
However, this scheme has several weaknesses, which limit it in practical applications. 
As shown in Fig.~\ref{fig:com_1}, their method requires multiple complex semantic segmentation network and feature extraction network to generate semantic maps and masks. Instead, we just design a simple
network (Masking) to identify the important regions of the image and generate the importance map to guide the allocation of bits.
Secondly, due to the complexity of their entire architecture, their codec efficiency is so low that it can't meet the needs of practical applications at all.
Moreover, changing the compression ratio has to reset parameters and retrain the model in their framework.

\begin{figure*}
	\begin{center}
		\includegraphics[width=1\linewidth]{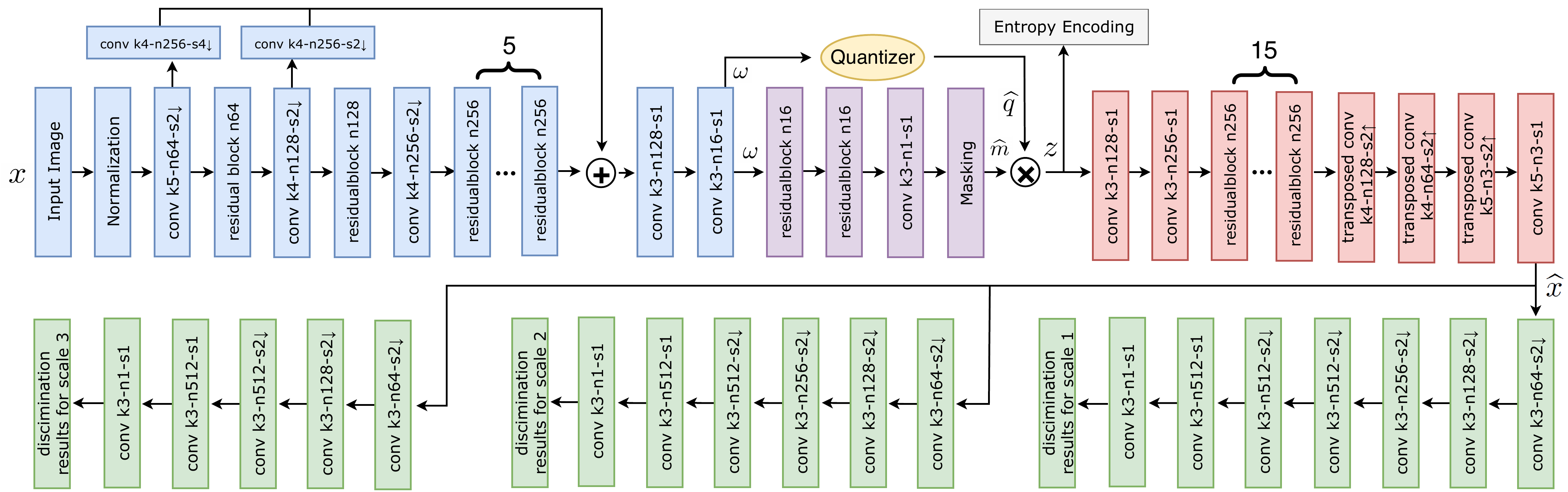}
	\end{center}
	\caption{Illustration of the GAN-based tunable image compression system. In the figure, blue, purple, yellow, gray, red, and green blocks represent encoder, masker, quantizer, encoder, decoder, and discriminator, respectively. It is noted that ``k5-n64-s2$\downarrow$'' represents a convolution layer with 64 filters of size 5 $\times$ 5 and a stride of 2. Each residual block has a uniform structure composed of two convolutional layers followed by a batch normalization \cite{ioffe2015batch} and a \textsl{ReLU} \cite{glorot2011deep}. Masking is an operation that extends the importance map to importance matrix according to Eq.(\ref{equation 3})}
	\label{fig:architecture}
\end{figure*}

\section{Model}\label{sec 3}

\subsection{Architecture}
Our image compression system is composed of six parts: encoder, quantizer, masker, entropy encoder, decoder, and discriminator. The entire architecture is shown in Fig.~\ref{fig:architecture}.
For a given image $x$ $\in$ X, the encoder converts it into a compact code matrix $\omega$ = $E(x)$, by multiscale convolution operations.
The masker takes $\omega$ as the input and generates an importance matrix $\widehat{m}$ = $M(\omega)$ through a simple convolutional network\cite{li2018learning,mentzer2018conditional} to guide the bit-allocation.
The quantizer $Q(\cdot)$ quantizes $\omega$ by using a nearest neighbor principle \cite{agustsson2017soft,mentzer2018conditional,theis2017lossy} and outputs $\widehat{q}$ = $Q(\omega)$.
The output of the quantizer and the masker are multiplied to gain the content-based image compression result, denoted as $z$ = $\widehat{m}$$\textbf{\ ·\ }$$\widehat{q}$.
The masker here can be understood as obscuring the non-important regions in the image and allocating more bits to the important regions.
Entropy encoder is applied to the system to remove the data redundancy and outputs $\widehat{h}$ = $H(z)$.
The decoder $G(\cdot)$, which is also named as generator, is the inverse of the encoder and generates the reconstructed image $\widehat{x}$ = $G(z)$.
The discriminator $D(\cdot)$ is an important part of the GAN, which improves the compression performance through alternating training \cite{agustsson2018generative} with the generator.

The six components of the image compression system will be introduced in the rest of the sections in detail.

\subsubsection*{Encoder}
\begin{figure}[h]
	\begin{center}
		\includegraphics[width=1\linewidth]{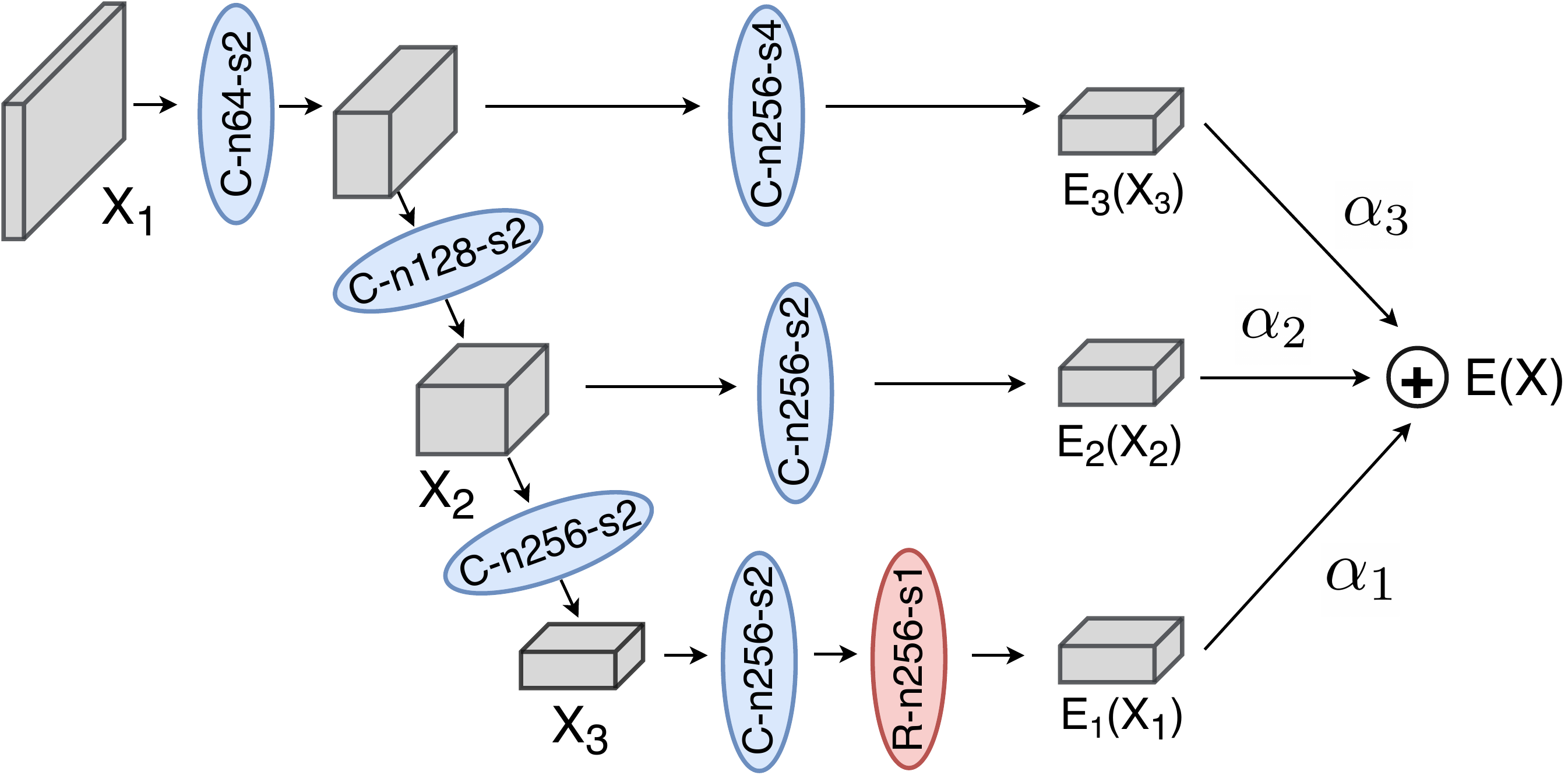}
	\end{center}
	\caption{Illustration of the encoder's pyramidal decomposition structure with 3 scales. It's noted that ``C-n16-s2'' represents a convolutional layer with 16 filters and a stride of 2 and ``R-n256-s1'' represents a residual block with 256 filters and a stride of 1.}
	\label{fig:pyra}
\end{figure}

\ \ \ \ In our image compression system, a fully convolutional neural network is used as the encoder, which consists of a crossover stack of several convolutional layers and residual blocks. 
To improve the compression performance of high-resolution images, we adopt the pyramidal decomposition scheme shown in Fig.~\ref{fig:pyra} to our model. 

Let $x_m$ denotes the input of the scale $m$ layer, so $x_1$ denotes the original input image. 
$E_m(x_m)$ represents the output of the scale $m$ layer. 
In our paper, we set $m$ to 1, 2, and 3 sequentially and execute encoding individually for each scale. 
The results of each scale are weighted and summed to produce an output $E(x)$ = $\alpha_1E_1(x_1)$ + $\alpha_2E_2(x_2)$ + $\alpha_3E_3(x_3)$.
Finally, $E(x)$ is convoluted with two convolutional layers to get the output of the encoder $\omega$ with the dimension of $\frac{H}{8}$$\times$$\frac{W}{8}$$\times$$K$. 
According to \cite{agustsson2018generative}, different $K$ produces different compression effects, which is a trade-off between the compression ratio and the distortion.

\subsubsection*{Masker}
\begin{figure}[htbp]
	\begin{center}
		\subfigure[]{ 
			\includegraphics[width=0.32\linewidth]{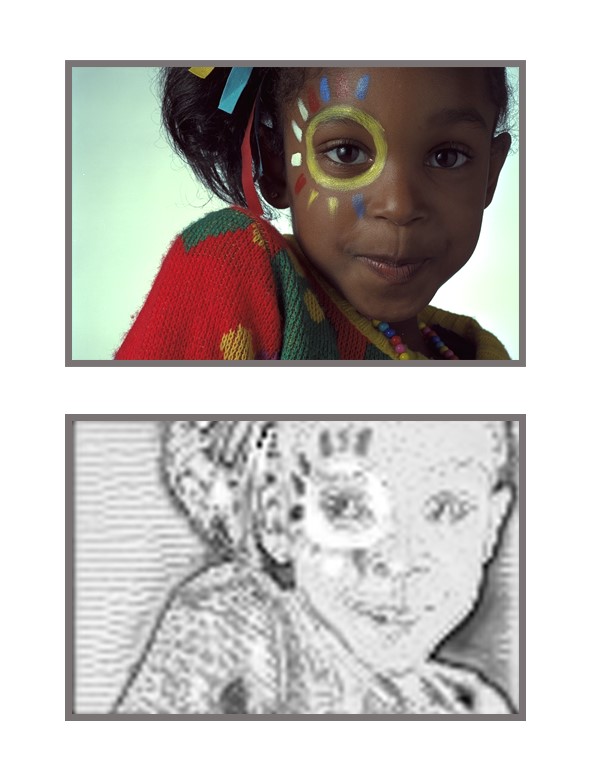}}
		\subfigure[]{ 
			\includegraphics[width=0.64\linewidth]{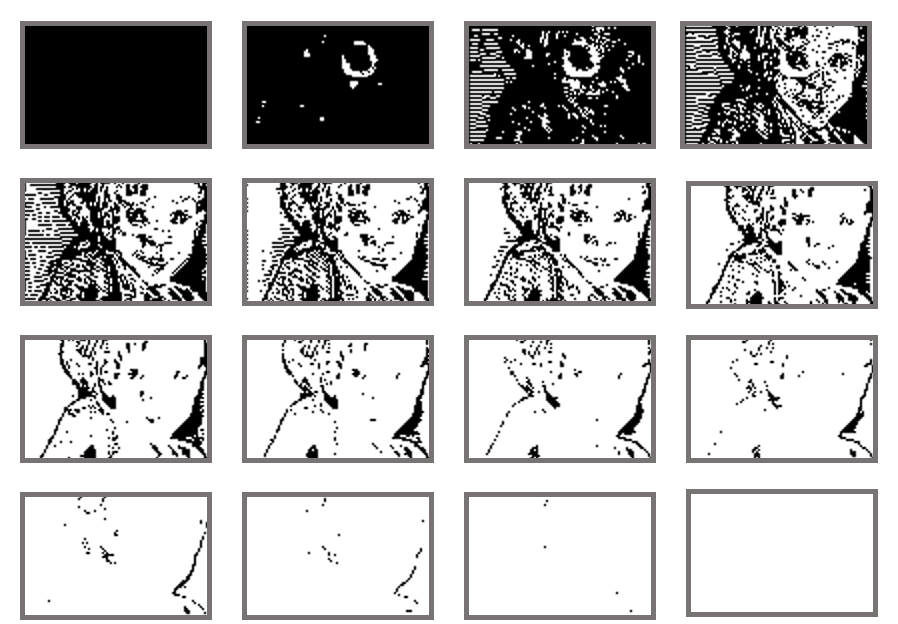}}
	\end{center}
	\caption{(a) The upper image is the original image $x$, and the lower image is the importance map $m$ \ (b) Images of each channel of the importance matrix $\widehat{m}$ (take $K$ = 16 ans $n$ = 0 as an example). The size of the $\widehat{m}$ is $\frac{H}{8}$$\times$$\frac{W}{8}$$\times$$K$. From left to right, top to bottom, the channel of the matrix $\widehat{m}$ gradually rises. The black regions represent importance regions.}
	\label{fig:imp}
\end{figure}

\ \ \ \ \ In an image, we tend to be interested only in some regions, which provides leeway for further improvement in compression ratio.
For example, for the portrait shown in Fig. \ref{fig:imp}, we are only interested in the face and body regions, which are called the important regions. 
The natural idea is that more bits are allocated to the important regions and fewer bits are allocated to the non-important regions.
The bit allocation according to the importance of image contents is achieved by constructing a masker.
%

\begin{figure*}[b]
	\centering
	\includegraphics[width=1\linewidth]{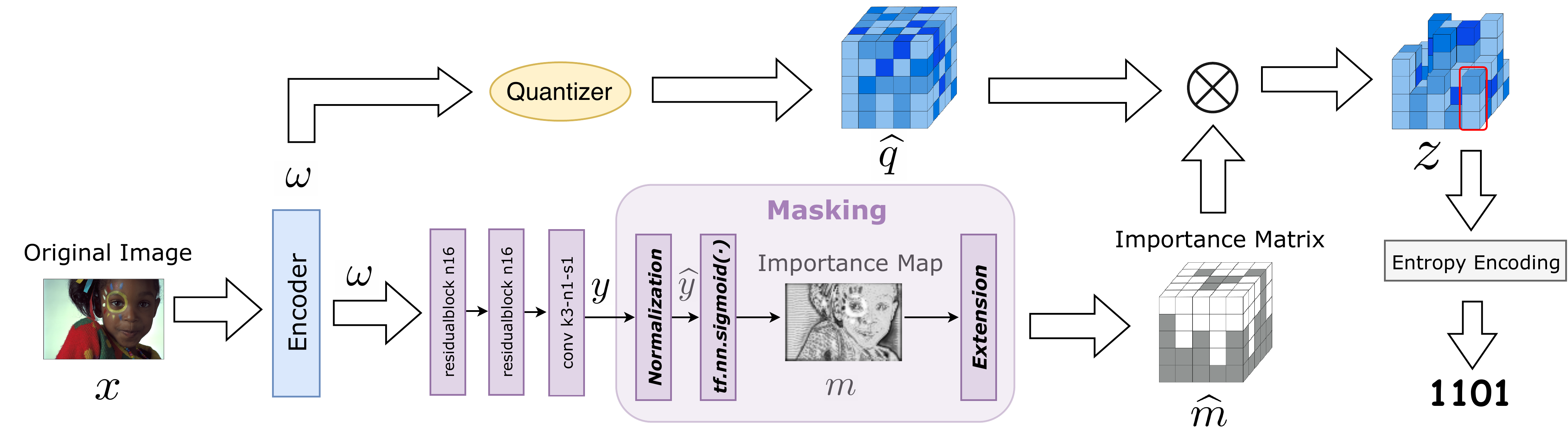}
	\caption{Illustration of how to generate an importance map $m$ and an importance matrix $\widehat{m}$ from the original image $x$ and use the importance matrix $\widehat{m}$ to guide the bit-allocation of the quantization result $\widehat{q}$. The colors of cubes in $\widehat{q}$ are from light to deep, respectively, and the quantized values are from 0 to 3. The gray and white cubes in $\widehat{m}$ represent 1 and 0, respectively. The light purple box corresponds to the \textbf{Masking} block in Fig.~\ref{fig:architecture}. \emph{Normalization}, \emph{tf.nn.sigmoid($\cdot$)} and \emph{Extension} correspond to the preprocessing of $y$ and the extension of the $m$. If the Huffman encoding is specified, that is 0-1, 1-01, 2-001 and 3-0001, then the circled part of $z$ is quantized to 1101.}
	\label{fig:importance}
\end{figure*}

The output of the encoder $\omega$ is used as the input of the masker, which is convoluted with two residual blocks.
Each residual block has 256 filters.
The kernel size and stride length of each filter are 3 $\times$ 3 and 1, respectively.
These residual blocks are followed by a convolutional layer with 1 filter.
The size of the output matrix $y$ is $\frac{H}{8}$$\times$$\frac{W}{8}$$\times$1. 
As shown in Fig.~\ref{fig:importance}, a \textsl{Sigmoid} activation is used to map the data $y$ to the range [0,1] to get an importance map $m$. 
However, the data after \textsl{Sigmoid} may converge to 0 or 1, resulting in the vanishing of the importance feature of $m$. To avoid this issue, we normalize the data in the matrix $y$ before the activation. The formula of the normalization procedure is specified as
\begin{equation}\label{equation 2}
\widehat{y}_{i,j}=\frac{y_{i,j}-\overline{\mu}}{\overline{\sigma}}\ ,\ m_{i,j}=tf.nn.sigmoid(\widehat{y}_{i,j})
\end{equation}

\noindent where $y_{i,j}$ represents the data value of the i-th row and the j-th column in the matrix $y$ and $\widehat{y}_{i,j}$ represents the data value of the i-th row and the j-th column in the matrix $\widehat{y}$. $\overline{\mu}$ and $\overline{\sigma}$ are the mean and variance of the matrix $y$, respectively. 
As mentioned earlier, we aim to design an tunable image compression system to compress the image to any $bpp$ without retraining the model. So here we replace $\overline{\mu}$ with $\overline{\mu}$ + $\emph{n}$ in the Eq.(\ref{equation 2}), \label{sec 3.1}where $\emph{n}$ is a random number within the range of [-2,2], and it is reassigned before each training batch.

The importance map $m$ is extended by the formula Eq.(\ref{equation 3}) to the importance matrix $\widehat{m}$, as shown in Fig.~\ref{fig:importance}.
\clearpage
\begin{equation}
\widehat{m}_{i,j,k} = \begin{cases}
0, & \mbox{if }\ \ m_{i,j} < \frac{k-1}{K} \\
1, & \mbox{if }\ \ m_{i,j} \ge \frac{k-1}{K}
\end{cases}
\ \ k = 1, \ldots, K
\label{equation 3}
\end{equation}
where $m_{i,j}$ represents the data value of the i-th row and the j-th column in the matrix $m$, and $\widehat{m}_{i,j,k}$ represents the data value of the i-th row, the j-th column and the k-th channel in the matrix $\widehat{m}$.
Taking the image $x$ in Fig.~\ref{fig:importance} as an example, the original image, the importance map and the images of each channel in the importance matrix are shown in Fig.~\ref{fig:imp}.
As the channel rises, the bits are mainly allocated to the face and body regions, and hardly in the background, which contributes to the improvement of the compression ratio. 
However, the image may be severely distorted at low $bpp$ due to the insufficient bit allocation in the background regions.
To address this issue, we reconstruct the non-important regions of the image based on GAN to improve the performance.

\subsubsection*{Quantizer} 
\ \ \ \ The selection of quantization bit is very important to the quantizer. 
Appropriate quantization bit not only improves the compression ratio, but also reduces the distortion.
We set $C_L$ = \{0, 1, 2, ..., $2^L$-1\}, and there are plenty of methods to quantize the input to a number in $C_L$. 
Here we use the nearest neighbor quantization method\cite{agustsson2017soft,mentzer2018conditional,theis2017lossy} to compute:
\begin{equation}
\widehat{q}=Q(\omega)=\arg\min\limits_{j}|\omega-c_j|
\end{equation}
where $c_j$ = $j$, and $j$ $\in$ $C_L$ = \{0, 1, 2, ..., $2^L$-1\}.

\subsubsection*{Entropy encoder} 
\ \ \ As shown in Fig.~\ref{fig:architecture}, in our method, images are reconstructed directly from the bitstream $z$ instead of the quantization results $\widehat{q}$ and the importance matrix $\widehat{m}$, which is exactly what we differ from other methods. 
The importance matrix $\widehat{m}$ indicates the non-mask bits (in gray) and the mask bits (in white) in $\widehat{q}$.
As shown in Fig~\ref{fig:importance}, the code matrix $z$ obtained by multiplying the results of masker and quantizer leaves only the non-mask bits.
Then we encode each channel of $z$ from the bottom up in a row-by-row manner.
Here we can only encode the non-mask bits. 
A large number of mask bits (data 0) at the top of each channel in $z$ can be encoded as a termination code instead of being encoded one by one.
When the mask bits are much more than the non-mask bits, our compression ratio can reach very low $bpp$. 
Since only a few non-mask bits at the bottom of each
channel need to be encoded, we use simple Huffman coding for entropy coding in this work\cite{han2015deep}.

\subsubsection*{Decoder} 
\ \ \ \ The decoder is the inverse of the encoder, and its function is to generate images with minimal distortion from the compression code matrix $z$. 
A good decoder should make restructured images as similar as possible to the original images in terms of texture, color, and so on.
The decoder is composed of a stack of 3 convolution layers, 15 residual blocks, and 3 transposed convolution layers. 
%
%
First of all, the input $z$ is convoluted with 128 filters of size 3$\times$3 and stride 1.
After that, the obtained feature maps are convoluted with 256 filters of size 3$\times$3 and stride 1, followed by 15 residual blocks. 
Similar to \cite{li2018learning}, these residual blocks are identical in the proposed model, consisting of two convolutional layers with 256 filters. 
%
%
Finally, the output of the last stage residual block is passed through 3 transposed convolutional layers to generate a reconstructed image $\widehat{x}$.

In fact, the decoder is also the generator in our GAN-based system. 
It improves its performance during the alternating training with the discriminator and generates images that the discriminator cannot identify the authenticity.

\subsubsection*{Discriminator}
\ \ \ \ The discriminator $D(\cdot)$ is able to identify the authenticity of the input image, i.e., whether it is the original image or the reconstructed image. 
As an important part of GAN, the discriminator $D(\cdot)$ is trained in parallel with the generator $G(\cdot)$\cite{agustsson2018generative,rippel2017real} to improve the performance of generating images.
In this paper, we continue to use the idea of pyramidal decomposition to design a multiscale discriminator. 
The motivation for adopting a multiscale architecture is to minimize the distortion at each scale separately.
For example, artifacts such as noise and blurriness are more easily found and eliminated at shallower scales, but the differences of the structure and the color of the image are usually found at deeper scales.
Here we assume that $\widehat{x}$ is the input of the discriminator $D(\cdot) $, and the input $\widehat{x}_m$ of the corresponding scale $m$ is obtained by the average pooling layer with a stride of 2.
The input of each scale passes through a convolutional network $D_m(\cdot)$ to produce an output $D_m(\widehat{x}_m)$.
Each convolutional layer of the networks is followed by \textsl{Leaky ReLU} instead of \textsl{ReLU} as the activation\cite{wang2018high}. 
%

\subsection{Loss function}\label{sec 3.2}
In the previous sections, we have designed an image compression system based on GAN.
Now we train the model on a batch of $B$,  that is $X_B$=\{$X^{(1)}$,$X^{(2)}$,$\cdots$,$X^{(B)}$\},\ containing high-resolution images.
The loss function of our model is composed of the following two parts.

\subsubsection*{Adversarial Loss}
\ \ \ We design our compression system based on GAN.
The generator $G(\cdot)$ is trained in parallel with the discriminator $D(\cdot)$. 
We call this part of the loss as adversarial loss, which is composed of the losses from generator $G(\cdot)$ and discriminator $D(\cdot)$. The adversarial loss is defined as follows:
\begin{equation}
\mathcal{L}_A = \sum_{i=1}^m\beta_i\Big\{\mathbb{E}[logD_i(x)]
+\mathbb{E}\big[log(1-D_i(G(x)))\big]\Big\}
\end{equation}
where $x$ is the original image, $m$ is the scale of discriminator, and $\beta_i$ is the weighting factor for scale $i$.

\subsubsection*{Distortion Loss}
\ \ \ \ The distortion loss measures the distortion of the original image $x$ and reconstructed image $\widehat{x}$. 
The purpose of the training is to minimize the following loss:
\begin{equation}
\mathcal{L}_D = E[d(x,\widehat{x})]
\end{equation}
where $d(\cdot)$ is a function to measure the similarity of the original image $x$ and reconstructed image $\widehat{x}$. 
In this paper, the Mean Square Error (MSE) is used in the distortion loss.

%
%

\subsubsection*{Overall Loss}
\ \ \ \ \ There is a constraint relationship between the above two losses. For example, increasing the adversarial loss may produce more generated contents in the reconstructed image, resulting in an increase in the distortion loss. 
Therefore, in the training process, we should consider the above two losses comprehensively. 
%
%
%
Since we train the model on the batch $X_B$ of size $B$, we need to consider the loss on the entire batch. 
The overall loss function is expressed as
\begin{equation*}
\begin{split}
\mathcal{L}_{G,D,E,B} &\!=\!\frac{1}{B}\sum_{j=1}^B\Big\langle \eta\sum_{i=1}^m\beta_i\Big\{\mathbb{E}\big[log(1\!-\!D_i(G(x^j)))\big]\\
&+\mathbb{E}[logD_i(x^j)]\Big\}+\kappa E\big[d(x^j,\widehat{x}^j)\big]\Big\rangle
\end{split} (9)
\end{equation*}
The training purpose is to minimize the overall loss.
\begin{equation}
\min\limits_{G,E,B}\min\limits_{D}\;\mathcal{L}_{G,D,E,B}
\end{equation}

\begin{figure*}[b]
	\centering
	\includegraphics[width=0.4975\linewidth]{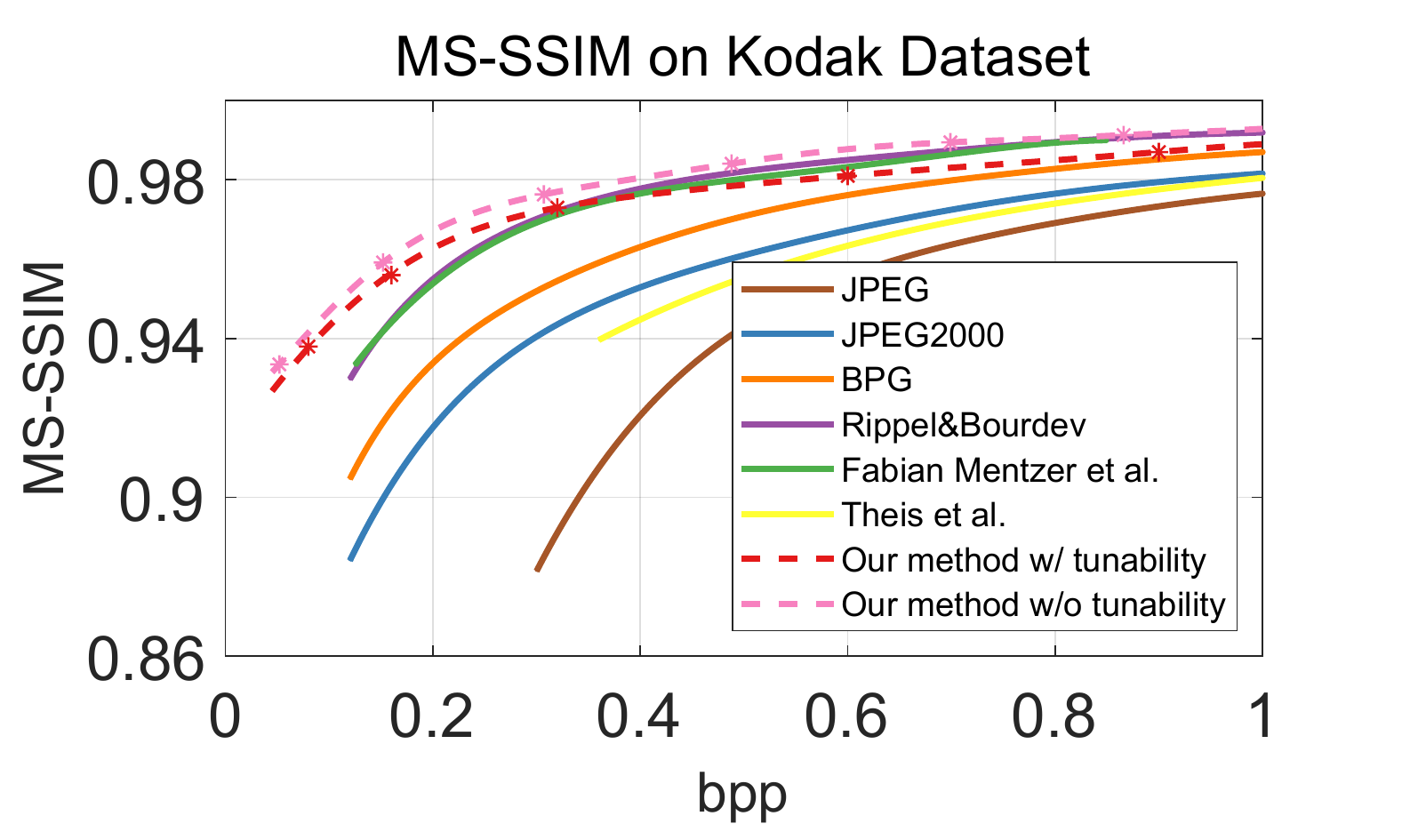}
	\includegraphics[width=0.4975\linewidth]{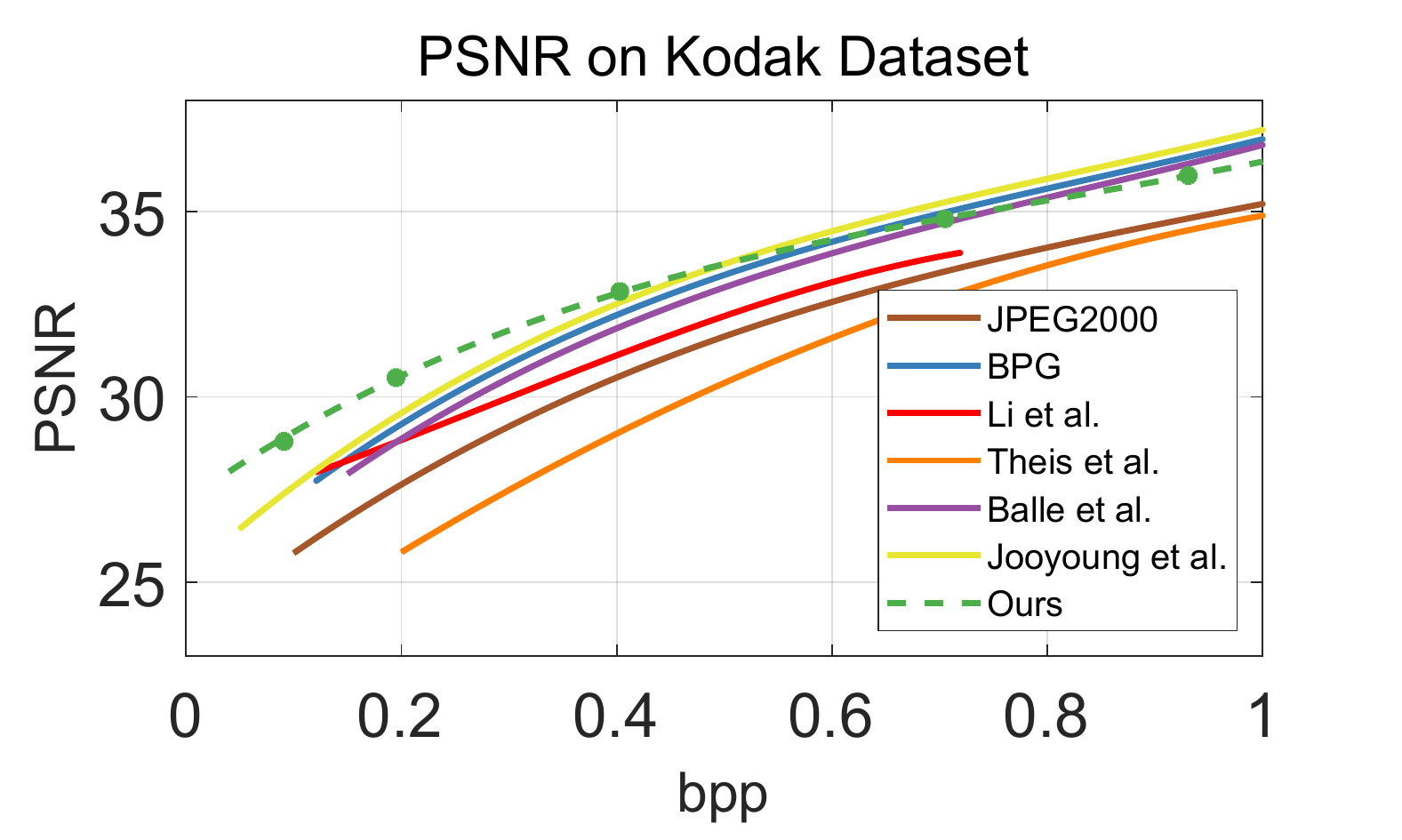}
	\caption{Comparison of compression performance by different methods measured by MS-SSIM and PSNR.}
	\label{fig:comparasion}
\end{figure*}

\section{Experiments and Results}\label{sec 4}
Our GAN-based tunable image compression system is trained on a subset of 15000 images in the ImageNet database\cite{russakovsky2015imagenet}. All images are scaled to 768 $\times$ 512, and every eight images are packed into a batch.
Then, we test the model on the Kodak dataset \cite{RN16}\, which is specifically designed to test the performance of lossy image compression. 
The compression ratio of the image is evaluated by $bpp$, which is the average number of bits required per pixel to store the compressed result. 
%
%
The distortion between the original and restructured image is commonly measured by MSE \cite{agustsson2018generative,agustsson2017soft,balle2016end,mentzer2018conditional,theis2017lossy}, PSNR, MS-SSIM\cite{rippel2017real,mentzer2018conditional,toderici2017full}. 
Compared with MSE, PSNR and MS-SSIM are used in this paper because they are more consistent with the actual perception of human vision\cite{patel2019human}.
%
In addition, we also compare the performance of our model on several different datasets.

In the rest of this section, we first introduce the parameter settings of our model, then compare the performance and visual effects of different compression methods and datasets. Besides, we perform an ablation experiment to show the impact of the importance matrix, entropy coding, GAN. Finally, we analyze the compression tunability of our system.

\subsection*{parameter settings}
Firstly, we set the weights $\alpha_1$, $\alpha_2$, and $\alpha_3$ of 3 scales in the encoder to 1/2, 1/4, and 1/4, respectively. 
Similarly, the weights $\beta_1$, $\beta_2$, and $\beta_3$ of 3 scales in the decoder are set to 1/2, 1/4, and 1/4, respectively.
The weights $\eta$ and $\kappa$ of two loss components of the overall loss are set to 1 and 16, respectively. 
In addition, we set the batch parameter $B$ to 8, which is to train the model using 8 images as a batch. Moreover, we set the quantization parameter $L$ to 2, which means $C_L$ = \{0, 1, 2, 3\}. 
%
%
In this work, if not specifically mentioned, let $K$ = 16. 
During the training process, the model is iteratively trained 128 times on the dataset. 
The initial learning rate is set to 2$\times10^{-3}$, and after 64 iterations, the learning rate is changed to 2$\times10^{-4}$.

\subsection*{Comparison of different methods}
Firstly, We compare the MS-SSIM performance of our tunable and non-tunable method with some conventional methods such as JPEG\cite{wallace1992jpeg}, JPEG2000\cite{skodras2001jpeg}, and BPG on the Kodak dataset\cite{RN16}.
We divide our compression system into two cases: tunable and non-tunable, which can be achieved by setting $n$ to a random value and a fixed value during the training process, respectively.
In addition, some DNN-based methods, such as \cite{rippel2017real,mentzer2018conditional,theis2017lossy}, are also included in the comparison. 
%
%
%
%

%
As shown in Fig.~\ref{fig:comparasion}, our method outperforms JPEG, JPEG 2000, BPG and the method proposed by Theis et al\cite{theis2017lossy} at a wide range of scale. 
At high $bpp$, the performance of our method is close to that of Mentzer et al.\cite{mentzer2018conditional} and Rippel \& Bourdev\cite{rippel2017real}, but at low $bpp$, our performance is much better than their methods.
For example, compared with the method proposed by Mentzer et al., the $bpp$ of our tunable and non-tunable models is reduced by 30.1\% and 39.3\% when MS-SSIM is 0.95, respectively. 

Next, we further compare the PSNR performance of different methods based on the work of \cite{lee2018context}. We compare our non-tunable method with JPEG 2000 and BPG as well as the methods proposed in \cite{li2018learning,theis2017lossy,balle2018variational}. In addition, the latest content-adaptive method proposed by Jooyoung et al. is also included in the comparison. As shown in Fig.~\ref{fig:comparasion}, at low $bpp$, the PSNR performance of our method is still
superior to other methods. At high $bpp$ ($bpp$ $\geqslant$ 0.5), our method will be slightly worse than \cite{lee2018context}, because at this time even the non-importance regions have been allocated enough bits, more and more contents generated by the GAN result in a decrease in PSNR performance. However, as we have always emphasized, our method focuses on the performance at extreme low $bpp$, and it’s acceptable to have a general performance at high $bpp$. 
The performance on different datasets is available in the supplementary materials.

\subsection*{Ablation experiments}
\begin{figure}[h]
	\centering
	\includegraphics[width=1\linewidth]{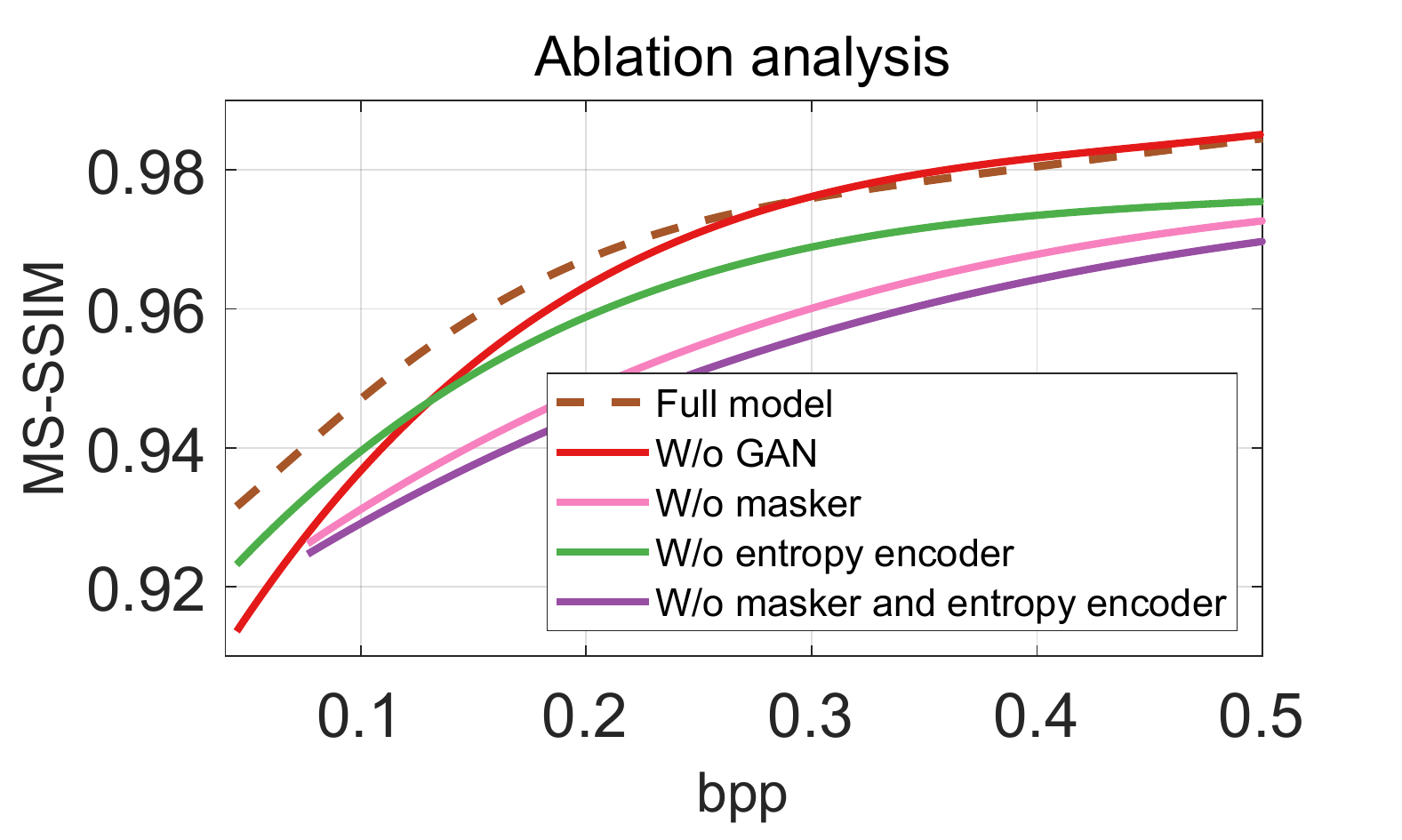}
	\caption{Illustration of the results of the ablation experiment.}
	\label{fig:ablation}
\end{figure}

The use of GAN in our proposed image compression system is to eliminate the distortion caused by insufficient bit allocation to non-important regions rather than generate new image contents.
In our method, sufficient bits are allocated to the important regions, which can be reconstructed realistically, and GAN has less impact on them. Under such circumstances, the main basis for the discriminator to discriminate is the non-important regions. So to confuse the discriminator, the generator will focus on reconstructing non-important regions.
The role of the importance matrix is to guide the allocation of bits and improve the representative efficiency of the bits\cite{agustsson2018generative}. 
The function of entropy coding is to further reduce data redundancy by exploiting the specificity of data distribution in the importance matrix.
We design the following five models according to whether the presence of  GAN, masker and entropy coding in the architecture: (1) full model; (2) model without GAN; (3) model without masker; (4) model without entropy coding; (5)model without masker and entropy coding; 
As shown in Fig.~\ref{fig:ablation}, under the same compression performance, (1) has the best performance while (5) has the worst performance. When MI-SSIM is 0.96, (2), (3), (4), and (5) has 15.8\%, 89.3\%, 31.6\%, and 116.3\% higher $bpp$ than (1), respectively.
The performance of the model with GAN performs better than that of the model without GAN at low $bpp$.
However, the improvement is gradually diminished with an increase of $bpp$. 
At high $bpp$, the introduction of GAN may even slightly impair compression performance. 
This shows that GAN's help with image compression is more pronounced at low $bpp$, which is not mentioned in other GAN-based methods.
In the compression task, we usually want the $bpp$ to be as small as possible, so the introduction of GAN can help solve the bottleneck limiting of the performance improvement at low $bpp$.

\subsection*{Tunability analysis}
Our system has the tunable characteristic, which means we can reassign the user-defined parameter $n$ in the masker to achieve different compression ratios without retraining the model. However, in the methods like\cite{li2018learning,mentzer2018conditional}, for an image, one type of network structure corresponds to one unique importance map. Therefore, only by modifying and retraining the model, it is possible to obtain different importance map, then achieve different compression ratios.

\begin{figure}[h]
	\centering
	\includegraphics[width=1\linewidth]{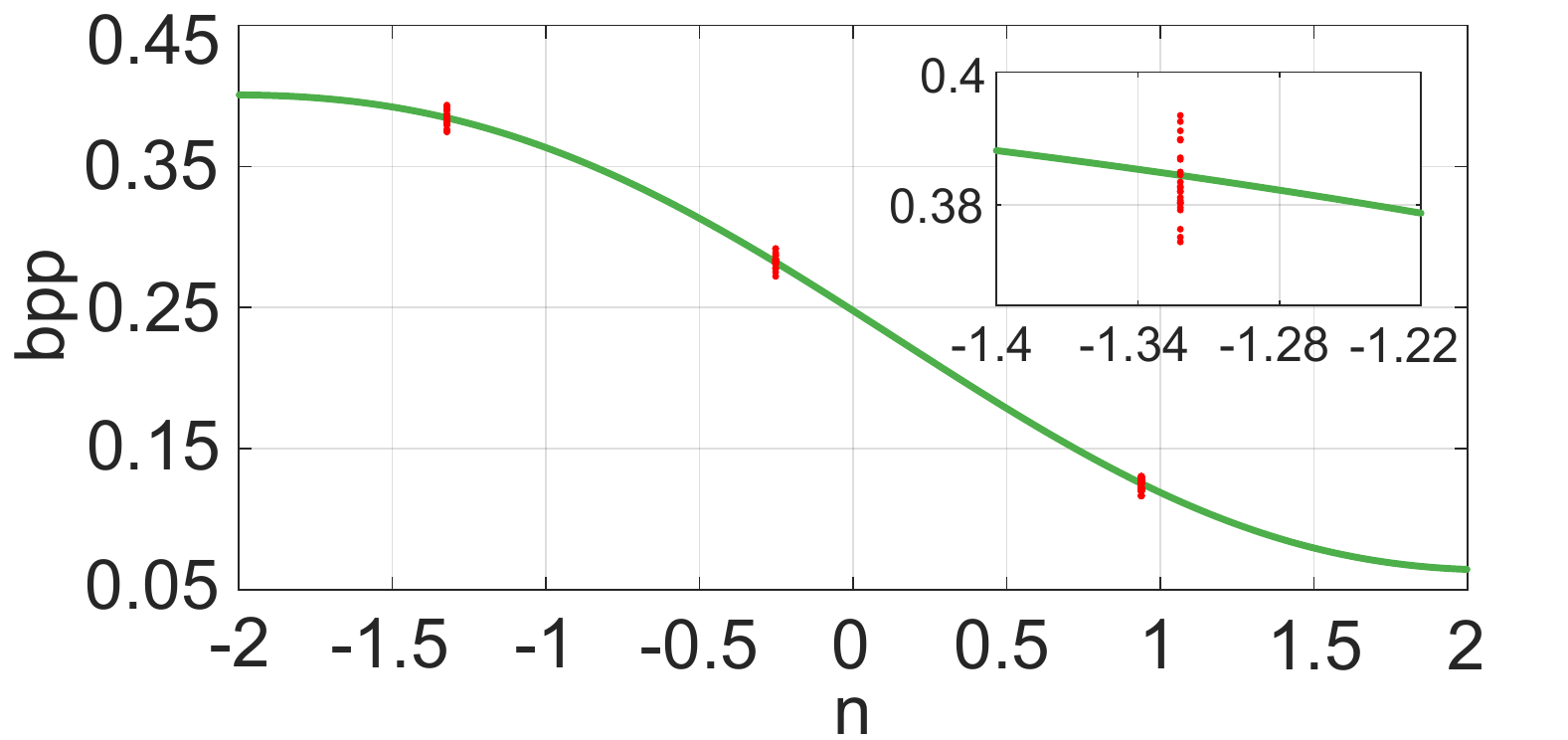}
	\caption{Tunability characteristic curve of the image compression system. The red dots represent the different compression ratios tested on the Kodak dataset at a fixed $n$.}
	\label{fig:interactive}
\end{figure}

The compression ratio of the image is determined by the parameter $\emph{n}$, which is an intuitive and simple dependency.
However, it should be noted that $\emph{n}$ and $bpp$ may not be in a strictly linear relationship. 
In the process of testing, different $\emph{n}$ is used to get its corresponding $bpp$. The data are fitted by Least Squares Method (LSM) to gain the tunability characteristic curve, as shown in Fig.~\ref{fig:interactive}. 
The image can be compressed to any specific $bpp$ within the range of [0.05, 0.4] as long as we set $\emph{n}$ to the corresponding value.
For example, if we want to compress an image to 0.384 $bpp$, then by looking up the figure and setting $n$ to -1.32, we can get a compression ratio around 0.384 $bpp$. The red dots in Fig.~\ref{fig:interactive} are the test results of images on the Kodak dataset.

\subsection*{Comparison of visual effects}
\begin{figure*}[t]
	\begin{center}
		\subfigure[Original]{ 
			\includegraphics[angle=90,width=0.115\linewidth]{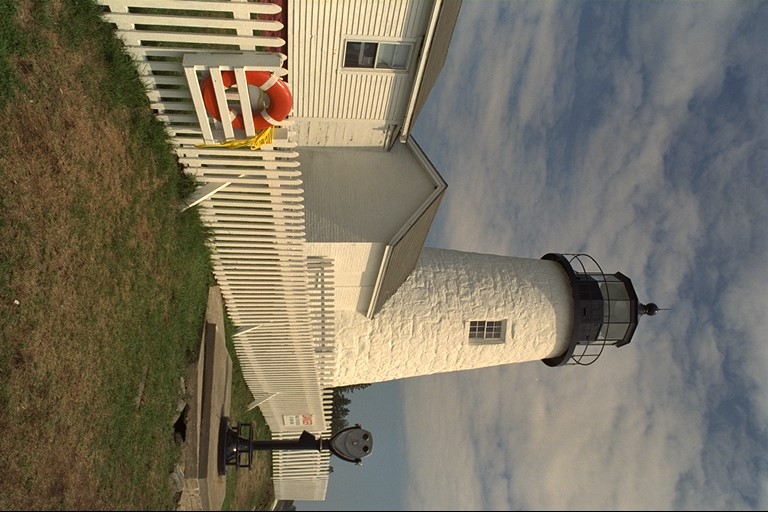}}
		\subfigure[Jpeg]{ 
			\includegraphics[angle=90,width=0.115\linewidth]{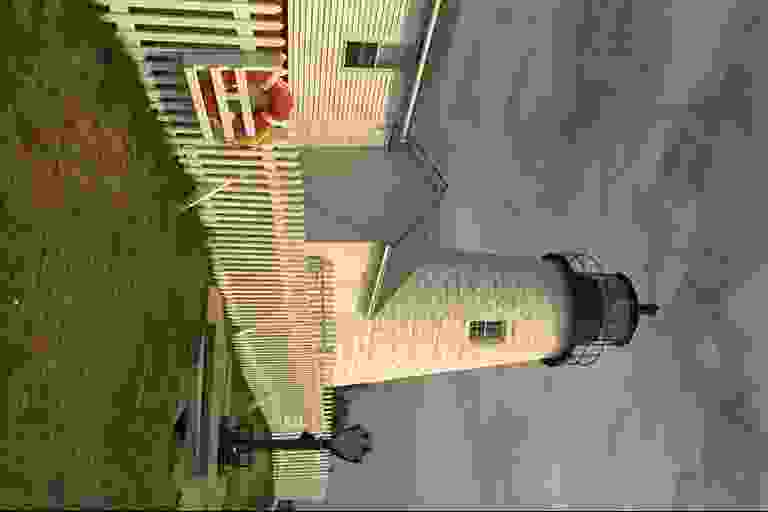}}
		\subfigure[Jpeg2000]{ 
			\includegraphics[width=0.115\linewidth]{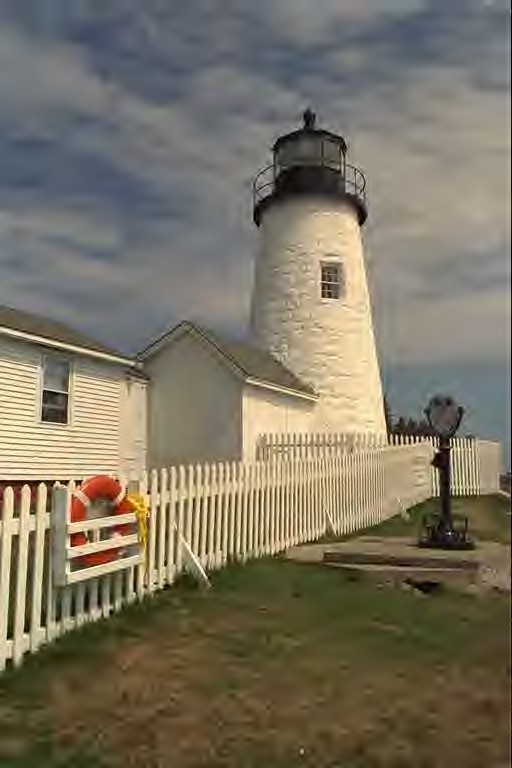}}
		\subfigure[Bpg]{ 
			\includegraphics[width=0.115\linewidth]{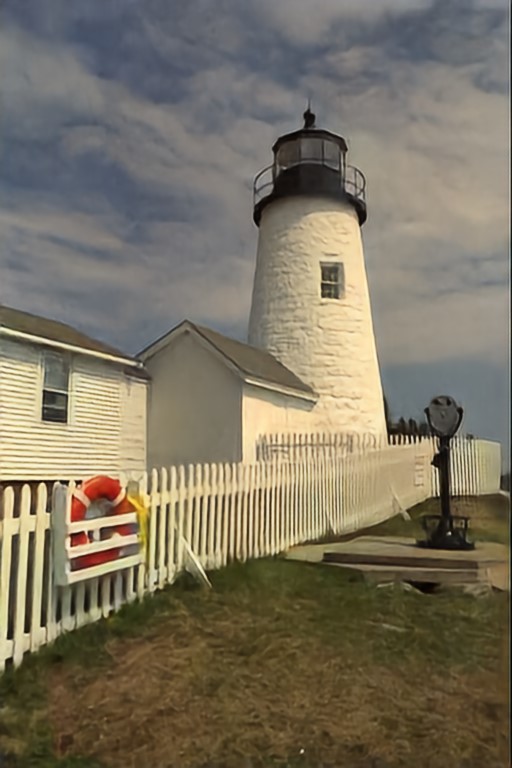}}
		\subfigure[Mentzer et al.]{ 
			\includegraphics[width=0.115\linewidth]{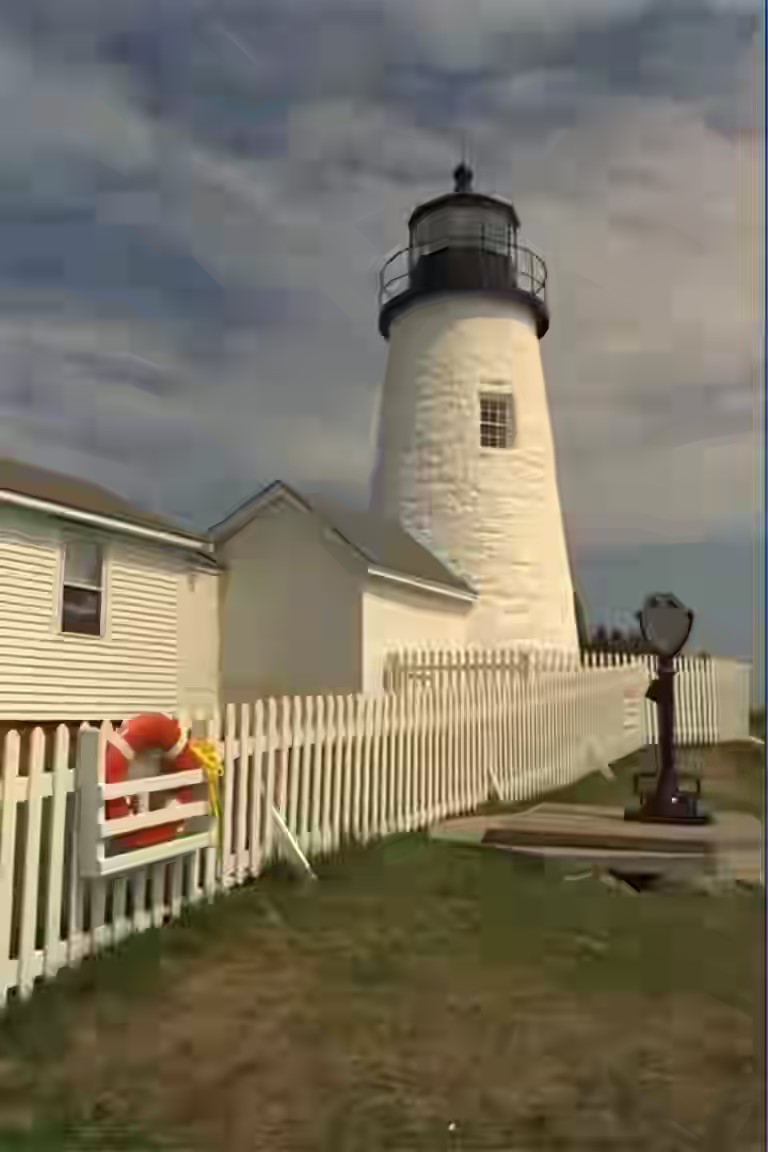}}
		\subfigure[Rippel et al.]{ 
			\includegraphics[width=0.11\linewidth]{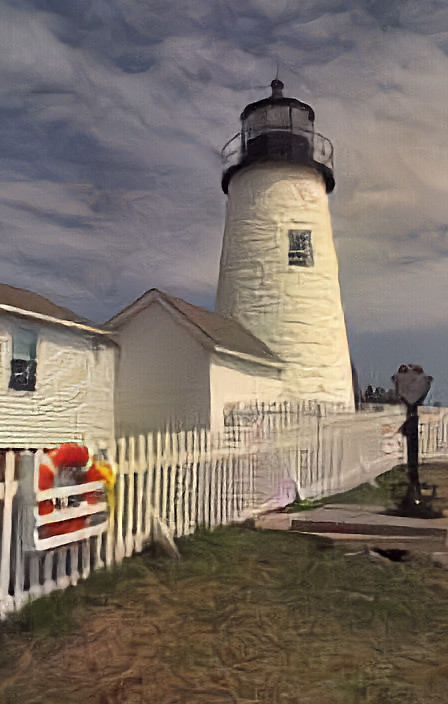}}
		\subfigure[Tunable]{ 
			\includegraphics[width=0.115\linewidth]{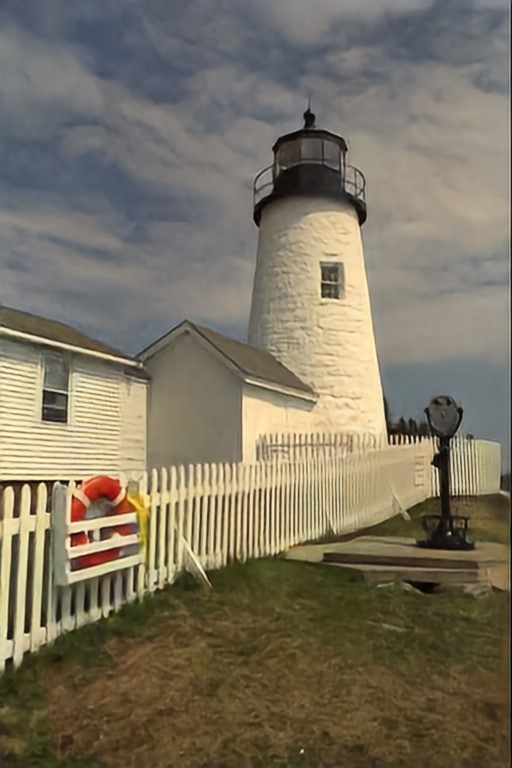}}
		\subfigure[Not tunable]{ 
			\includegraphics[width=0.115\linewidth]{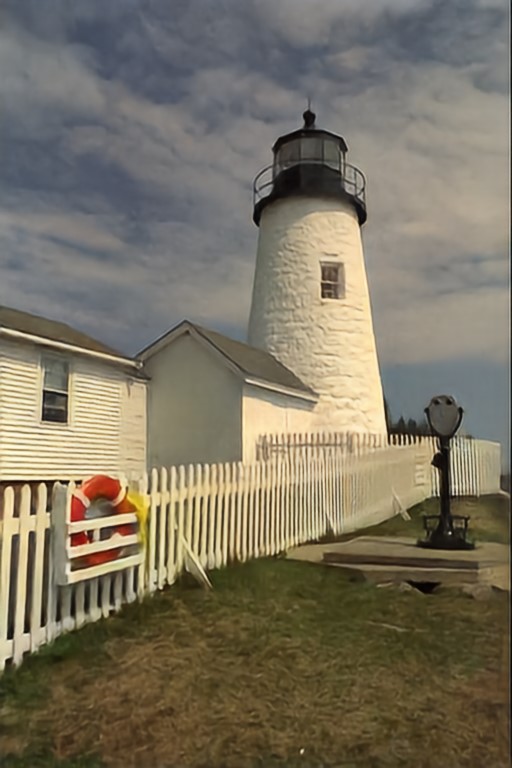}}  
	\end{center}
	\caption{Illustration of the original image and the reconstructed images produced by conventional and DNN-based compression methods. From left to right, the $bpp$ of each method is 24 $bpp$, 0.123 $bpp$, 0.125 $bpp$, 0.108 $bpp$, 0.128 $bpp$, 0.093 $bpp$, 0.116 $bpp$, 0.103 $bpp$.}
	\label{fig:result_2}
\end{figure*}

In Fig.~\ref{fig:result_2}, we compare our methods with JPEG, JPEG 2000, BPG as well as the methods of Fabian Mentzer et al.\cite{mentzer2018conditional} and Rippel \& Bourdev\cite{rippel2017real} visually. 
As can be seen from Fig.~\ref{fig:result_2}, conventional image compression methods such as JPEG, JPEG 2000 and BPG inevitably produce blurring, ringing, etc.\cite{agustsson2018generative}, which can seriously affect the human visual experience. 
Though the methods of Fabian Mentzer et al.\cite{mentzer2018conditional} and Rippel \& Bourdev\cite{rippel2017real} are very good at detail processing, they fail to show the structure and color of the image well. 
In contrast, our method overcomes the above flaws, and some important colors and textures are well-retained and more visually pleasing due to the bit-allocation based on the image contents.

\begin{figure*}[t]
	\begin{center}
		\subfigure[$bpp$ / MS-SSIM: Ours 0.039 / 0.927, Agustsson et al. 0.030 / 0.824]{ 
			\includegraphics[width=0.16\linewidth]{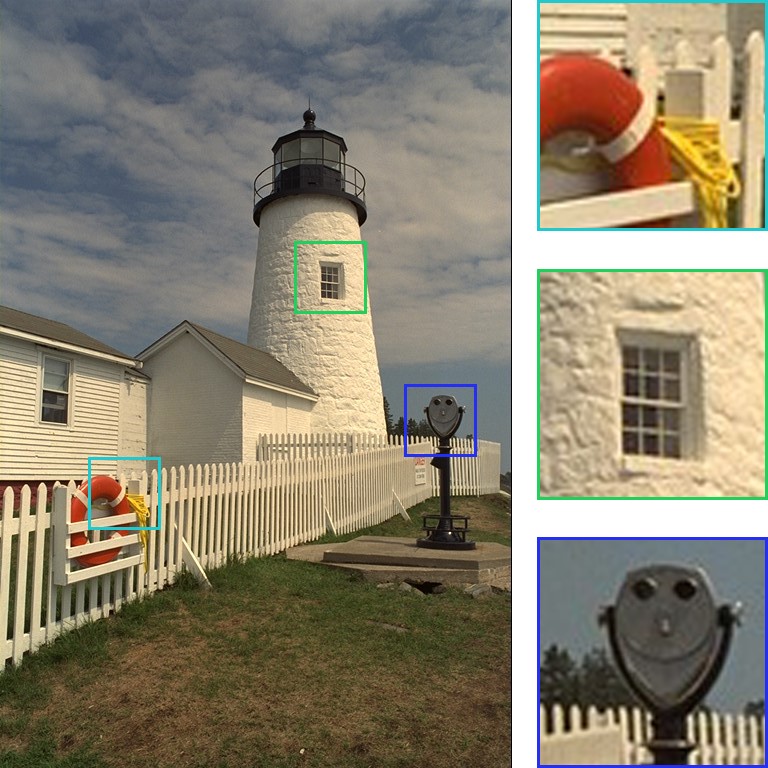}
			\includegraphics[width=0.16\linewidth]{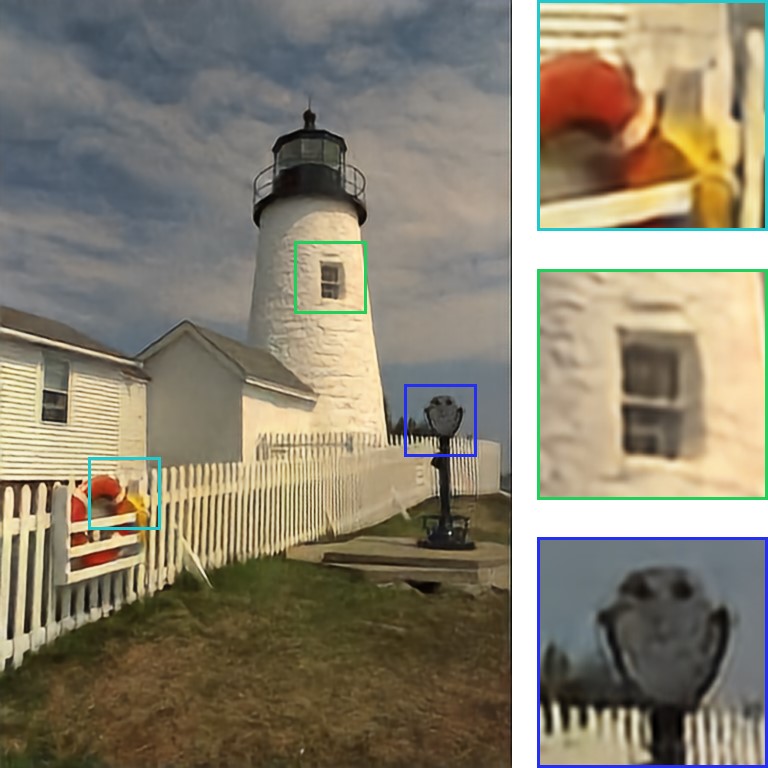}
			\includegraphics[width=0.16\linewidth]{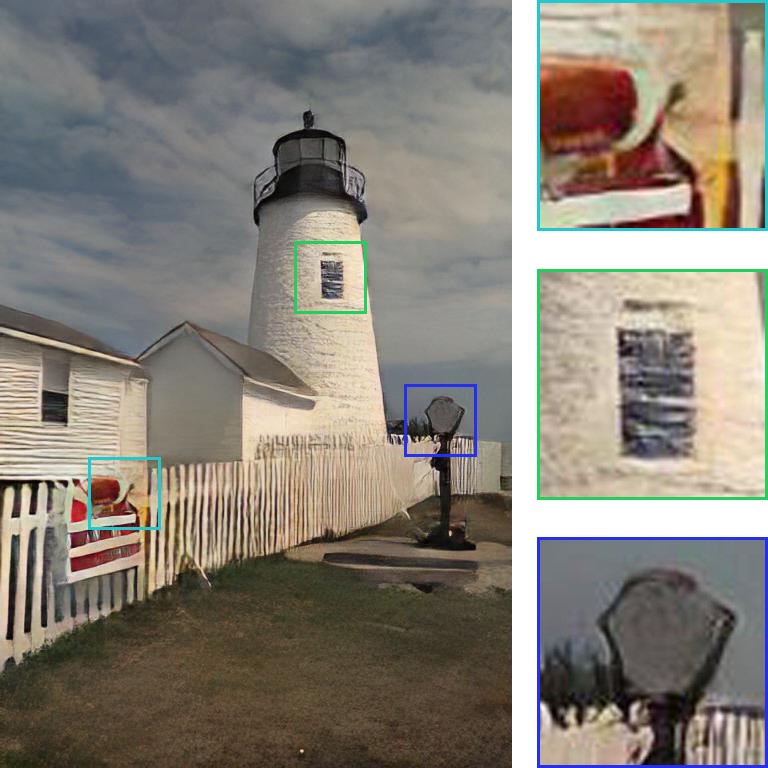}}
		\subfigure[$bpp$ / MS-SSIM: Ours 0.063 / 0.906, Agustsson et al. 0.069 / 0.795]{ 
			\includegraphics[width=0.16\linewidth]{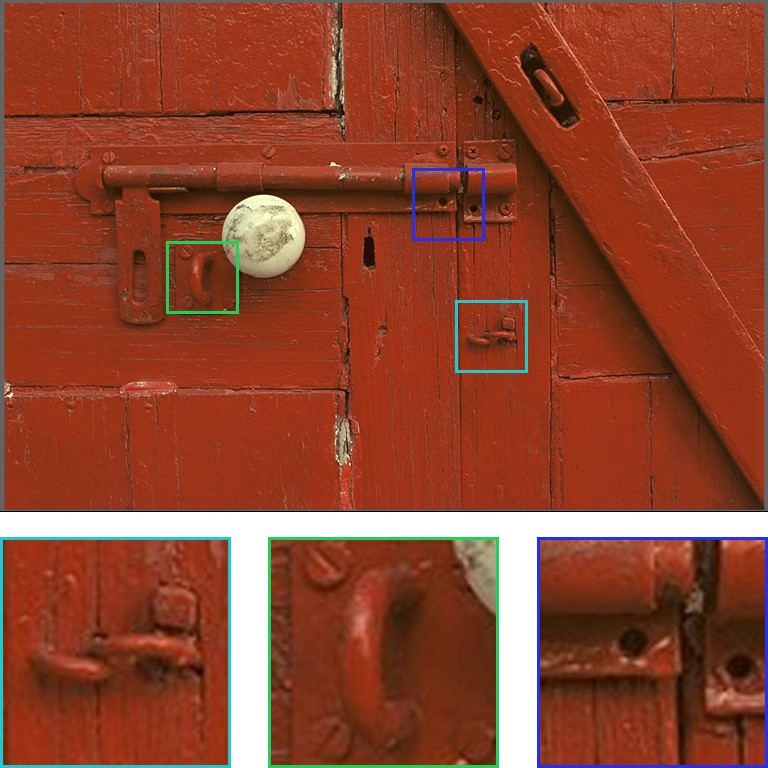}
			\includegraphics[width=0.16\linewidth]{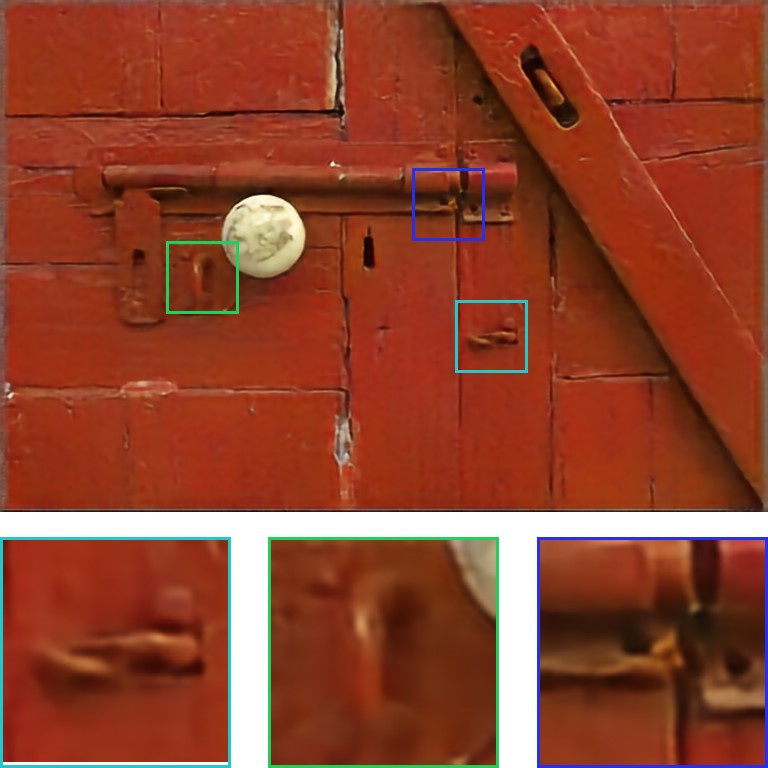}
			\includegraphics[width=0.16\linewidth]{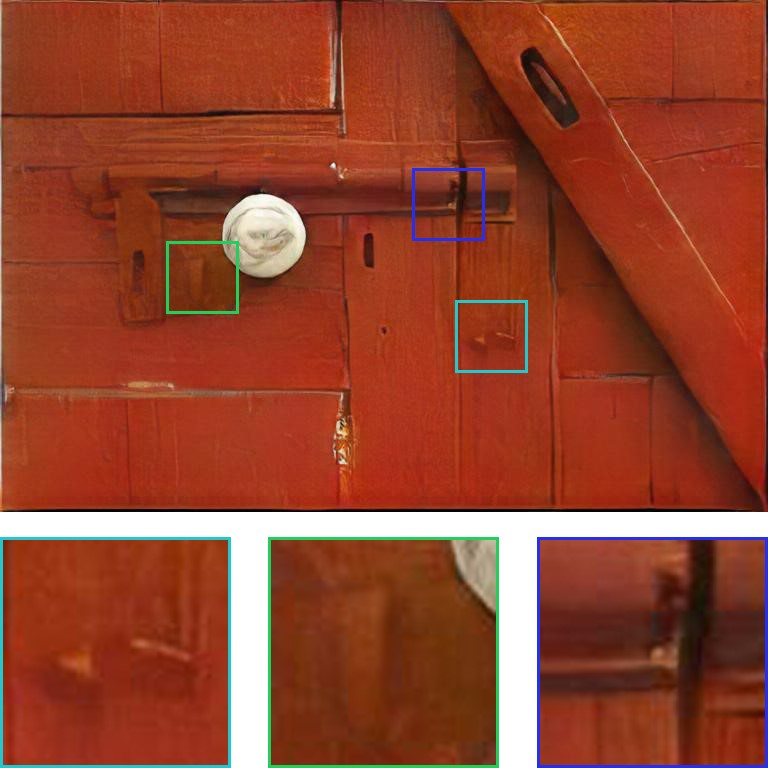}}
		\subfigure[$bpp$ / MS-SSIM: Ours 0.058 / 0.921, Agustsson et al. 0.065 / 0.845]{ 
			\includegraphics[width=0.16\linewidth]{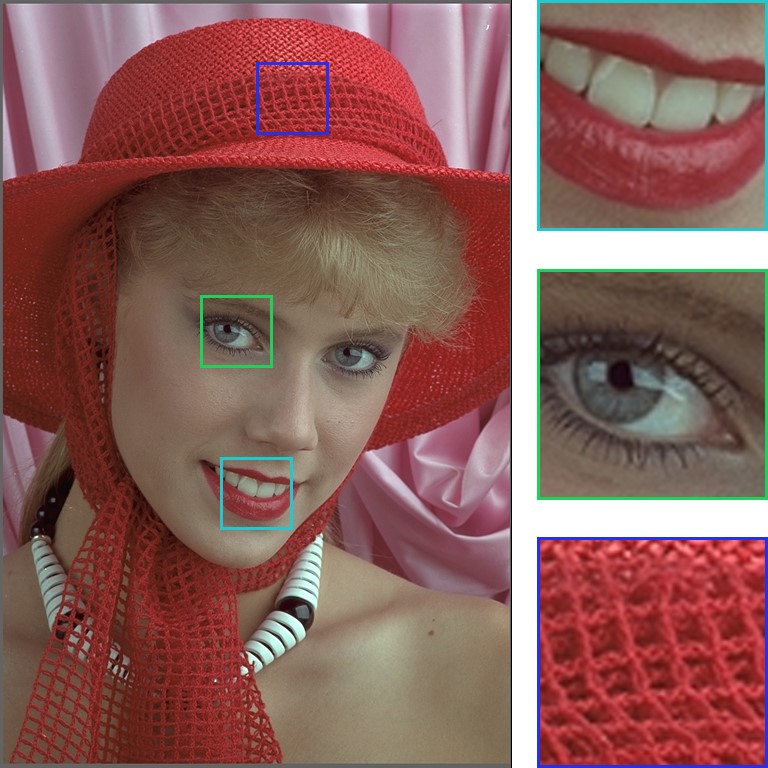}
			\includegraphics[width=0.16\linewidth]{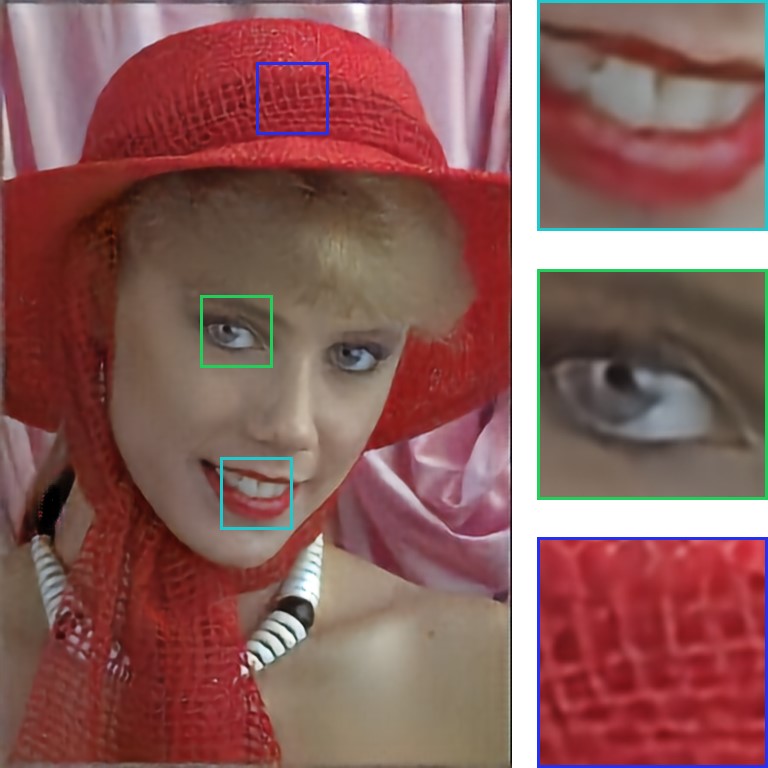}
			\includegraphics[width=0.16\linewidth]{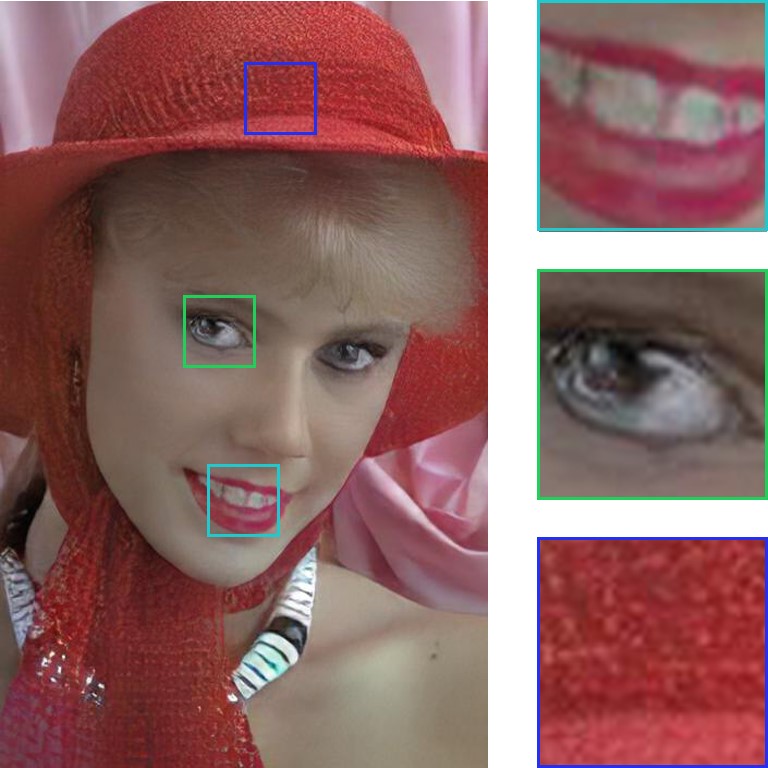}}  
		\subfigure[$bpp$ / MS-SSIM: Ours 0.040 / 0.937, Agustsson et al. 0.034 / 0.844]{ 
			\includegraphics[width=0.16\linewidth]{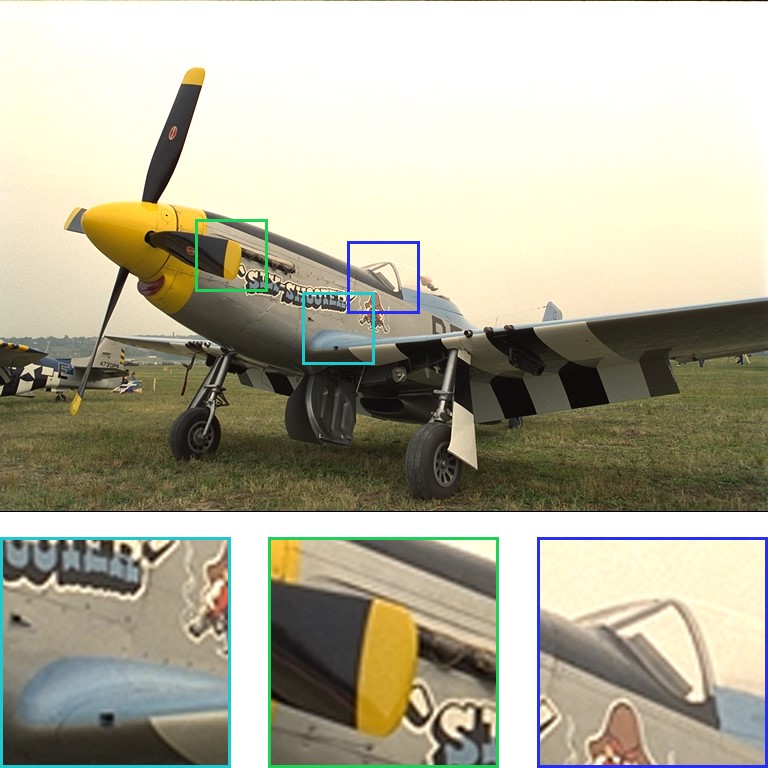}
			\includegraphics[width=0.16\linewidth]{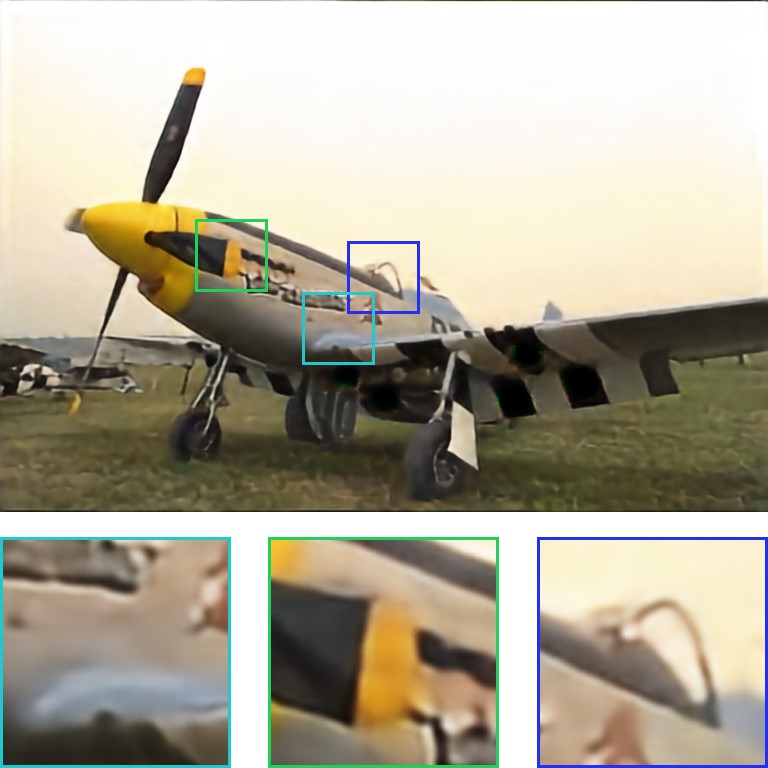}
			\includegraphics[width=0.16\linewidth]{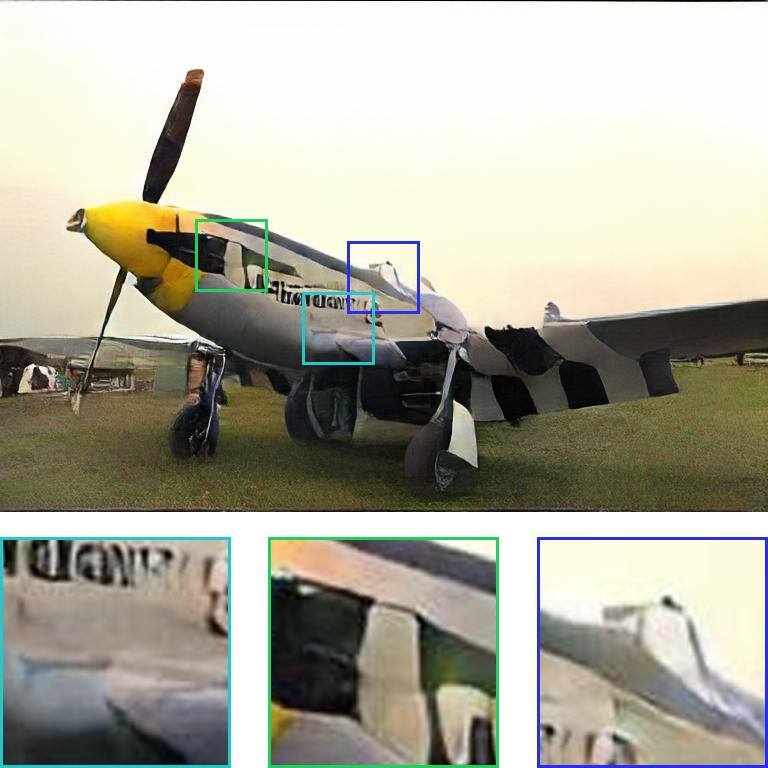}}  
	\end{center}
	\caption{Illustration of comparison with the state of the art GAN-based method. From left to right: Original, Ours, Agustsson et al.}
	\label{fig:result_3}
\end{figure*}

In Fig.~\ref{fig:result_3}, we compare our non-tunable method with the most advanced GAN-based method at low $bpp$. Compared with\cite{agustsson2018generative}, since we introduce the important matrix into our system, the details of the image, such as the window of the house, the lock on the door, the holes in the woman's hat and the fuselage and paddles of the aircraft, are well preserved.
Besides, due to the use of GAN, the non-importance regions of the image are also very harmonious, without severe distortion resulted from the lack of bits.
In term of MS-SSIM, since \cite{agustsson2018generative} is too dependent on GAN, their MS-SSIM is only 83.9\% at 0.05 $bpp$.
In contrast, our MS-SSIM is 10.3\% higher than theirs.
For more visual comparisons, please refer to the supplementary materials.

\section{Discussion}\label{sec 5}
In our architecture, we design the multiscale encoder and discriminator based on the idea of pyramidal decomposition, introduce importance map for bit allocation, and further compress data by entropy coding. 
As for the training approach, we introduce two losses, all of which are weighted and summed to get the overall loss function and use global compression of high-resolution images instead of block compression. 
At the same time, we introduce GAN to reconstruct non-important regions of the image to solve the distortion caused by insufficient bit allocation to non-important regions.

The experimental results show that our method outperforms the-state-of-art content-based and GAN-based methods when $bpp$ is smaller than 0.2.
At low $bpp$, the existence of GAN has a more significant impact on performance improvement because the insufficient bit allocation in the non-importance regions often occurs in the case of low $bpp$.
In terms of MS-SSIM and PSNR, our method is superior to conventional compression algorithms such as JPEG, JPEG2000 and BPG, and also outperforms the state-of-the-art DNN-based compression methods at low $bpp$.
Visually, our approach solves the flaws of conventional algorithms such as ringing, blurring, etc., and can better preserve the texture, color, and other details of the image.
In addition, as shown in Fig.~\ref{fig:interactive}, our system has the tunable characteristic, and within a certain range, the compression ratio of any $bpp$ can be achieved through an user-defined parameter $n$ without retraining the model. 
On the Kodak dataset, to achieve MI-SSIM of 0.95, the average time to encode and decode the image is 21 $ms$ and 29 $ms$, running on the GeForce GTX 1080 Ti.


\section{Conclusions}\label{sec 6}
In this paper, we have proposed a GAN-based tunable lossy image compression system.
In the proposed system, GAN is trained to reconstruct the non-important regions of the image and thus reduce the distortion caused by the insufficient bit allocation to those non-important regions. 
More importantly, the idea of tunability has been applied to the DNN-based image compression systems. Our compression system has the tunability characteristic, which means we can compress an image to a specific compression ratio without retraining the model.

\clearpage
\label{reference}
{\small
\bibliographystyle{ieee}
\bibliography{egbib}
}

\end{document}